# Tracing a Faint Fingerprint of the Invisible Hand?
## Retrieving the Progressive from the Recorded Movement of the Monetary Economy

Koichiro Matsuno

*Department of BioEngineering, Nagaoka University of Technology, Nagaoka 940-2188, Japan*
*Fax: +81 258479420;*kmatsuno@vos.nagaokaut.ac.jp; http://bio.nagaokaut.ac.jp/~matsuno

**Abstract.** Any economic agent constituting the monetary economy maintains the activity of monetary flow equilibration for fulfilling the condition of monetary flow continuity in the record, except at the central bank. At the same time, monetary flow equilibration at one economic agent constantly induces at other agents in the economy further flow disequilibrium to be eliminated subsequently. We propose the rate of monetary flow disequilibration as a figure measuring the progressive movement of the economy. The rate of disequilibration was read out of both the Japanese and the United States monetary economy recorded over the last fifty years.

**Keywords:** disequilibration, equilibration, invisible hand, monetary economy, monetary flow

## INTRODUCTION

The flow of funds accounts compiled and released by the monetary authority presents an intriguing historical record demonstrating how the monetary economy has developed as being driven by the dynamic motive force from within. Although it is not explicit in identifying the nature of the dynamic motive force, the record of the flow of funds accounts is arguably a historical witness to how the invisible hand after Adam Smith could have come to manipulate or regulate the monetary economy. We shall report our attempt for reading some faint trace of the dynamic motive force from the recorded flow of funds accounts released from both the Bank of Japan (BOJ) and the Federal Reserve Board (FRB) of the United States.

Our method of reading the trace of the dynamic motive force is simple and straightforward [1]. Imagine, for instance, that a wageworker happens to lose the current job and to decrease the monthly income because of the forced change of the income solely to the unemployment benefit. The unemployed worker is then forced to change the expenditure behavior in order to meet the balance of monetary flow between the reduced income and the intended expenditure, that is the activity of monetary flow equilibration. The unemployed worker first comes to experience the imbalance or disequilibrium of monetary flow momentarily and then to commit himself to monetary flow equilibration immediately afterward. Since the change in the expenditure behavior of the unemployed worker is then going to affect the sales of a nearby grocery store, monetary flow equilibration at the worker subsequently induces monetary flow disequilibrium at the grocery store. Monetary flow disequilibrium in the latter in turn necessarily comes to induce further activity of monetary flow equilibration since no one can counterfeit bank notes except the central bank.

Monetary flow disequilibrium thus reverberates in the monetary economy in a ceaseless manner. Although every economic agent constantly acts for eliminating the disequilibrium between the incoming and the outgoing monetary flow, monetary flow disequilibrium does not disappear altogether from the monetary economy because of the local nature of each bilateral transaction.

We measure the rate of monetary flow disequilibration as consulting the recorded flow of funds accounts for both Japan and the United States over the last fifty years.

## MONETARY DYNAMICS

For simplicity, we shall consider the monetary economy consisting of five aggregated agents; Corporations, Households, Financial Institutions, Government and Central Bank [1,2]. Monetary inflow to Corporations $\dot{y}_1^{(in)}$ consists of several subcategories as

$$\dot{y}_1^{(in)} = \dot{y}_{21} + \dot{y}_{31}^{(\ell)} + \dot{y}_{31}^{(s)} + \dot{y}_{41} \tag{1}$$

in which $\dot{y}_{21}$ is Households' payment for purchasing the commodities produced at Corporations, $\dot{y}_{31}^{(\ell)}$ is the loan to Corporations from Financial Institutions, $\dot{y}_{31}^{(s)}$ is the interest payment from Financial Institutions to the time deposits and savings made by Corporations and $\dot{y}_{41}$ is Government's payment for purchasing the commodities produced at Corporations. The dot appeared in variable $\dot{y}$ in the above does not and will not imply time derivative in what follows. It is just an integral part of the variable so designated. Monetary outflow from Corporations $\dot{y}_1^{(out)}$ consists of

$$\dot{y}_1^{(out)} = \dot{y}_{12} + \dot{y}_{13}^{(\ell)} + \dot{y}_{13}^{(s)} + \dot{y}_{14} \tag{2}$$

in which $\dot{y}_{12}$ is the wage payment to Households from Corporations, $\dot{y}_{13}^{(\ell)}$ is the interest payment by Corporations to the loan from Financial Institutions, $\dot{y}_{13}^{(s)}$ is the time deposits and savings at Financial Institutions made by Corporations, and $\dot{y}_{14}$ is the tax payment to Government by Corporations.

Monetary inflow to Households $\dot{y}_2^{(in)}$ is

$$\dot{y}_2^{(in)} = \dot{y}_{12} + \dot{y}_{32}^{(s)} + \dot{y}_{32}^{(\ell)} + \dot{y}_{42}^{(x)} + \dot{y}_{42}^{(nx)} \tag{3}$$

in which $\dot{y}_{32}^{(s)}$ is the interest payment by Financial Institutions to the time deposits and savings made by Households, $\dot{y}_{32}^{(\ell)}$ is the loan from Financial Institutions, $\dot{y}_{42}^{(x)}$ is the taxable wage payment to Households from Government, and $\dot{y}_{42}^{(nx)}$ is the nontaxable transfer income to Households processed by Government. Monetary outflow from Households $\dot{y}_2^{(out)}$ is

$$\dot{y}_2^{(out)} = \dot{y}_{21} + \dot{y}_{23}^{(s)} + \dot{y}_{23}^{(\ell)} + \dot{y}_{24} \tag{4}$$

in which $\dot{y}_{23}^{(s)}$ is the time deposits and savings at Financial Institutions made by Households, $\dot{y}_{23}^{(\ell)}$ is the interest payment by Households to the loan from Financial Institutions, and $\dot{y}_{24}$ is the tax payment to Government by Households.

Monetary inflow to Financial Institutions $\dot{y}_3^{(in)}$ is

$$\dot{y}_3^{(in)} = \dot{y}_{13}^{(s)} + \dot{y}_{23}^{(s)} + \dot{y}_{13}^{(\ell)} + \dot{y}_{23}^{(\ell)} + \dot{y}_{53}^{(i)} + \dot{y}_{43}^{(b)} + \dot{y}_{53}^{(b)} \tag{5}$$

in which $\dot{y}_{53}^{(i)}$ is the loan from Central Bank to Financial Institutions, $\dot{y}_{43}^{(b)}$ is the interest payment by Government to Governmental bonds possessed by Financial Institutions, and $\dot{y}_{53}^{(b)}$ is due to the buying operation of Governmental bonds by Central Bank. Monetary outflow from Financial Institutions $\dot{y}_3^{(out)}$ is

$$\dot{y}_3^{(out)} = \dot{y}_{31}^{(s)} + \dot{y}_{32}^{(s)} + \dot{y}_{31}^{(\ell)} + \dot{y}_{32}^{(\ell)} + \dot{y}_{34} + \dot{y}_{35}^{(i)} + \dot{y}_{35}^{(b)} + \dot{y}_{35} \tag{6}$$

in which $\dot{y}_{34}$ is the tax payment to Government by Financial Institutions, $\dot{y}_{35}^{(i)}$ is the interest payment to the loan from Central Bank, $\dot{y}_{35}^{(b)}$ is due to the selling operation of Governmental bonds by Central Bank, and $\dot{y}_{35}$ is the reserve deposit made at Central Bank by Financial Institutions.

Monetary inflow to Government $\dot{y}_4^{(in)}$ is

$$\dot{y}_4^{(in)} = \dot{y}_{14} + \dot{y}_{24} + \dot{y}_{34} + \dot{y}_{34}^{(b)} + \dot{y}_{54}^{(b)} \tag{7}$$

in which $\dot{y}_{34}^{(b)}$ is due to the payment by Financial Institutions for purchasing Governmental bonds from Government, and $\dot{y}_{54}^{(b)}$ is due to the payment by Central Bank for purchasing Governmental bonds from

Government. Monetary outflow from Government $\dot{y}_4^{(out)}$ is

$$\dot{y}_4^{(out)} = \dot{y}_{41} + \dot{y}_{42}^{(x)} + \dot{y}_{42}^{(nx)} + \dot{y}_{43}^{(b)} \tag{8}$$

Since no economic agent except Central Bank can counterfeit or destroy bank notes, the continuity of monetary flow

$$\dot{y}_j^{(in)} = \dot{y}_j^{(out)} \qquad (j = 1,2,3,4) \tag{9}$$

has to hold in the completed record compiled by an external observer, say, the monetary authority. Internally, however, monetary flow disequilibration

$$\Delta \dot{y}_j \equiv \dot{y}_j^{(in)} - \dot{y}_j^{(out)} \tag{10}$$

does not vanish altogether in a synchronous manner because there is no global means for total synchronization to be applied to all of the concerned parties exactly at the same instant [3].

When monetary flow disequilibrium happens to occur at Corporations as

$$\Delta \dot{y}_1 \neq 0 \tag{11}$$

at a certain instant, Corporations have to do every possible means for eliminating the disequilibrium at the immediately following instant. Otherwise, monetary flow continuity to be observed in the completed record by all means would be destroyed. However, the consequence of monetary flow equilibration at Corporations subsequently causes monetary flow disequilibration at Households, Financial Institutions and Government. It comes to appear through, for instance, decrease (increase) of the wage payment to Households, decrease (increase) of the loan from Financial Institutions, and decrease (increase) of the price of the commodities purchased by Government and Households. Monetary flow equilibration at one economic agent for the sake of monetary flow continuity in the record constantly serves as an agency for monetary flow disequilibration at other agents. A more detailed discussion on decision making is found elsewhere [1].

Monetary flow disequilibrium $\Delta \dot{y}_j(t)$ (j=1,2,3,4) at each agent at a given instant t is immediately subject to monetary flow equilibration by the same agent. The consequence of each act however does not fail to induce further monetary flow disequilibrium $\Delta \dot{y}_j(t + \tau)$ to be eliminated at the immediately following instant $t + \tau$, in which $\tau$ is the time interval for updating the decisions on the part of the participating agents. The updating process is summarized as

$$\Delta \dot{y}_j(t + \tau) = \sum_{k=1}^{4} B_{jk} \Delta \dot{y}_k(t) \tag{12}$$

in which $\{B_{jk}\}$ is the transfer matrix representing the activity of monetary flow disequilibration. While it actualizes monetary flow continuity right in the middle of the transfer process, each agent comes to suffer monetary flow disequilibrium after experiencing the concurrent acts of monetary flow equilibration at the other agents. Monetary flow disequilibrium $\Delta \dot{y}_j(t)$ cannot be frozen in the completed record. Instead, the transfer matrix $\{B_{jk}\}$ is real in the sense of leaving monetary flow continuity to be read out from the record as such by the external observer, e.g., the overseeing monetary authority, while carrying with itself monetary flow disequilibration forward.

One can in fact measure the development of the transfer matrix as consulting the pattern of decision making that all of the concerned agents update successively. Among many alternatives, a most significant figure characterizing the transfer matrix B is its maximum eigenvalue measuring the resultant rate of monetary flow disequilibration, that is, the relative growth rate of monetary flow disequilibrium per unit time for the monetary economy as a whole.

## OBSERVED RATE OF DISEQUILIBRATION

One can determine the transfer matrix of monetary flow disequilibration as consulting the record of the flow of funds accounts. The time series of the observed nine variables we referred to include the total outstanding of currency issued by Central Bank, the total outstanding of Governmental bonds issued by Government, the total outstanding of Governmental bonds possessed by Central Bank, the loan to Financial Institutions from Central Bank, the reserve deposit at Central Bank made by Financial Institutions, the discount rate set by Central Bank, the interest rate to Governmental bonds, the interest rate to the loan from Financial Institutions, and the interest rate to time deposits and savings at Financial Institutions.

We ran monetary dynamics of flow equilibration and disequilibration, and tried to simulate the observed time series of these nine different variables as closely as possible upon random search [1]. Even the time interval $\tau$ for updating each decision was taken as a parameter to optimally be determined through the random search. The time series of the nine variables were retrieved from the records of the flow of funds accounts compiled and released from both BOJ and FRB at every ten days starting from January 1950 through December 1999 [4,5]. When the first hand data at every ten days were not available, the linear interpolation was employed just for the purpose of data acquisition.

The rate of monetary flow disequilibration measured in units of percents per day was equated to and read from the maximum eigenvalue of the transfer matrix of monetary flow disequilibration. The typical sample trajectories of the rate of disequilibration for the Japanese and the United States monetary economy are presented in Figs. 1 and 2. A unique feature common to all of the sample trajectories we have tried so far is that there were sporadic bursts in the rate of monetary flow disequilibration, though lasted only over less than one month in every such occasion. The bursts occurred for Japan in early 1955, late 1959, 1968, late 1973 and mid 1987, and for the United States in early 1970, 1972, late 1974, 1989 and 1994. One functional role of bursting monetary flow disequilibration is to eventually regulate a growing monetary flow disequilibrium in a drastic manner from time to time.

The rate of disequilibration represents the strength of the dynamic motive force driving the monetary economy in the forward direction. Monetary flow disequilibrium experienced by one agent does induce the activity of monetary flow equilibration there immediately afterward, while it does not fail to subsequently cause monetary flow disequilibration at the other agents. When the rate of disequilibration remains positive, monetary flow disequilibration grows and reverberates in the economy. The persistent presence of monetary flow disequilibrium in the economy demonstrated for both Japan and the United States over the last fifty years manifests that the dynamic force identified as monetary flow equilibration for one party and at the same time as disequilibration for the others has constantly been operating. There have however been a few exceptional occasions such that the rate of disequilibration remained negative though only temporarily as demonstrated in early 1968, late 1976, and mid 1982 for Japan. The negative rate of disequilibration implies losing the dynamic motive force driving the monetary economy.

The significance of negative rate of disequilibration should also be noted in the light that monetary flow disequilibrium cannot grow indefinitely. Since no surviving economic agent is allowed to maintain its outstanding of currency negative at any time, the intensity of monetary flow disequilibration has to be regulated so as to keep the outstanding of currency possessed by any surviving agent always positive. The intensity of monetary flow disequilibration would certainly be curbed when the rate of disequilibration becomes negative. However, it could happen that the intensity of monetary flow disequilibration be overly regulated by major policy changes. When the monetary economy of Japan was subject to the BOJ's decisions of sudden hike of the discount rate from 2.50 to 3.75 percents per year within less than one year in 1989, the aftereffect may have reverberated in the economy at the least during the succeeding several years.

Historical analysis of the observed rate of disequilibration must however be equivocal in the best sense of the word, depending upon the theoretical premise and the analytical tool to be employed. Rather, because of its least vulnerability to prior theoretical commitments, the rate of disequilibration can be utilized as a tool for planning fiscal, monetary and investment policy on the part of the major players participating in the monetary economy. Its rationale rests upon the fact that the rate of disequilibration, though retrievable from the finished record, carries the capacity of driving the monetary economy forward.

Moreover, the monetary dynamics upon monetary flow equilibration and disequilibration can provide us with a means of anticipating to some extent how the monetary economy would develop in the near future. Since monetary flow equilibration as an organizing factor of the monetary economy also serves as a disturbing factor of the economy through the accompanying disequilibration, the actual economy must be robust enough against the disturbing factor. Otherwise, the actual development must have been replaced by other alternatives that could be more robust. The robustness of the economy against monetary flow disequilibration does provide a reliable reference against which an arbitrarily anticipated trajectory of the movement of the economy may be examined and evaluated with regard to its actual likelihood.

## CONCLUSIONS

Addressing the monetary economy is an ambivalent endeavor. Monetary dynamics of the economy certainly proceeds in a dialogic mode because of the participation of more than one agent, while the description of the dynamics has to be attempted in a monologic mode [6,7]. At issue is how to accommodate the dialogic movement of economic agents to the practice of monologic description. In this regard, the rate of monetary flow disequilibration we have elaborated serves as a unique means for bridging between the dynamic dialogue and the descriptive monologue. The rate of disequilibration addresses an actualization of monetary flow continuity accessible to the monologic description in between the preceding and the immediately following monetary flow disequilibration, the latter of which is accessible only to the participating economic agents. Precisely for this reason, the rate of disequilibration stands as a descriptive indicator of the dynamic motive force driving the monetary economy in the forward direction.

The way of interpreting the rate of disequilibration however can be different among fiscal policy maker, monetary authority, investor and consumer. Their primary objectives may be different. Nonetheless, one thing is shared by all of the concerned parties. That is, no one can take the rate of disequilibration to remain either positive or negative over an indefinite period of time.

## ACKNOWLEDGMENTS

Thanks are due to Nobuaki Ono, Kaoru Sugimoto, Kazuhiro Tanaka and Mitsuhiro Tomoda for their collaboration during the early phase of this work, to Koichi Hirano for compiling the economic data, and to Kuniyuki Hatori for help in the programming.

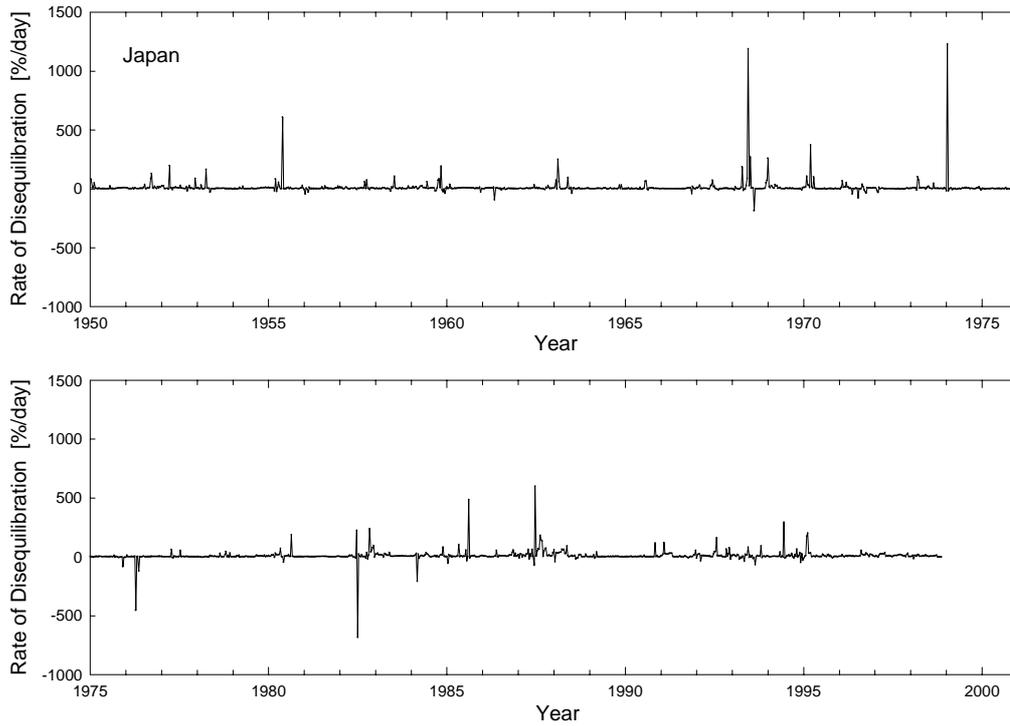

**FIGURE 1.** A typical sample trajectory of the rate of monetary flow disequilibration for the monetary economy of Japan over the period of 1950-1999. The rate of disequilibration measures the relative growth rate of monetary flow disequilibrium in units of percents per day.

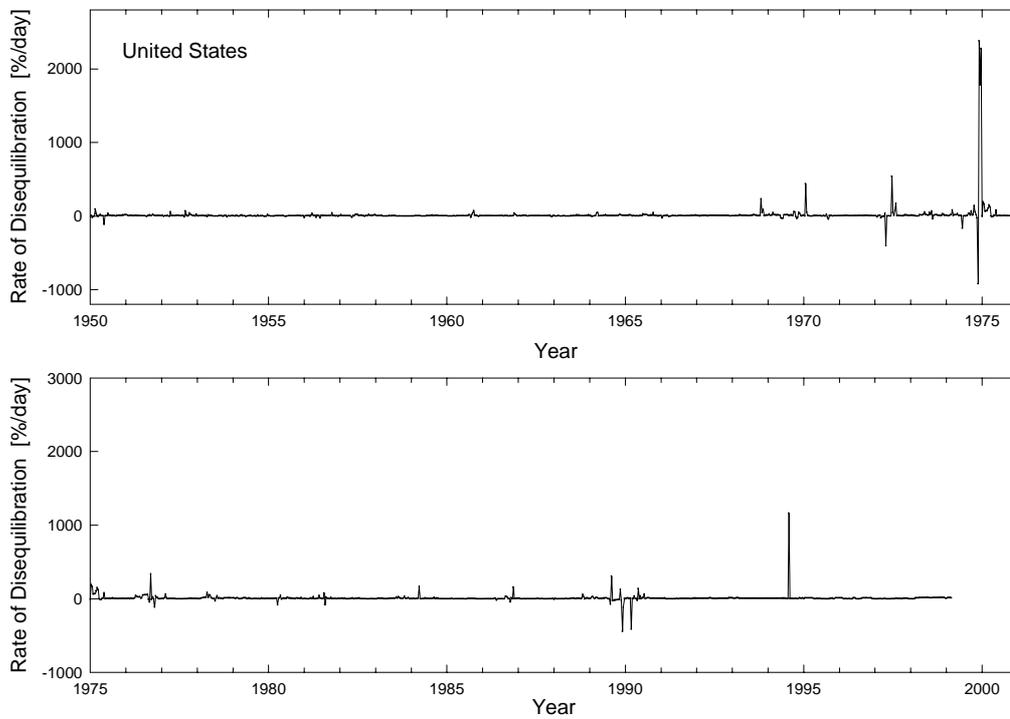

**FIGURE 2.** A typical sample trajectory of the rate of monetary flow disequilibration for the monetary economy of the United States over the period of 1950-1999. The rate of disequilibration measures the relative growth rate of monetary flow disequilibrium in units of percents per day.